\documentclass[iop,apj,twocolappendix]{emulateapj}

\usepackage{natbib}
\usepackage{amsmath,amssymb}
\usepackage{apjfonts}

\usepackage[usenames]{xcolor}
\definecolor{mpl_blue}{HTML}{1F77B4}
\definecolor{mpl_orange}{HTML}{FF7F0E}
\definecolor{mpl_green}{HTML}{2CA02C}
\definecolor{mpl_red}{HTML}{D62728}

\renewcommand{\check}{\textcolor{orange}}

\usepackage[backref, breaklinks, plainpages=false, colorlinks=true, anchorcolor=cyan, linkcolor=mpl_red, citecolor=cyan, urlcolor=magenta, bookmarks=false]{hyperref}
\usepackage[all]{hypcap}
\usepackage[caption=false]{subfig}

\shorttitle{NANOGrav $11$-year GW Memory}
\shortauthors{The NANOGrav Collaboration}

\newcommand{\Msun}{M_\odot}
\newcommand{\dd}{\mathrm{d}}
\newcommand{\psrAnomalyA}{J1909$-$3744}
\newcommand{\psrAnomalyB}{J0030$+$0451}
\newcommand{\psrBF}{J1744$-$1134}
\newcommand{\psrBestA}{J1713$+$0747}
\newcommand{\psrBestB}{J2317$+$1439}
\newcommand{\psrBestC}{J1600$-$3053}

\newcommand{\bayesephem}{\textit{BayesEphem}}

\newcommand{\hmem}{h_\mathrm{mem}}
\newcommand{\hopt}{h_\mathrm{opt}}
\newcommand{\fyr}{\mathrm{yr}^{-1}}

\defcitealias{ng5}{NG5}%{NG5a}
\defcitealias{ng5.bwm}{NG5mem}%{NG5b}
\defcitealias{ng9}{NG9}
\defcitealias{ng11}{NG11}%{NG11a}
\defcitealias{ng11.gwb}{NG11gwb}%{NG11b}
\defcitealias{ng11.cw}{NG11cw}%{NG11c}

\begin{document}

\title{The NANOGrav 11-Year Data Set: Limits on Gravitational Wave Memory}
\author{
%The NANOGrav Collaboration:
K.~Aggarwal\altaffilmark{1,2},
Z.~Arzoumanian\altaffilmark{3},
P.~T.~Baker\altaffilmark{${\color{magenta}\S}$4,2},
A.~Brazier\altaffilmark{5,6},
P.~R.~Brook\altaffilmark{1,2},
S.~Burke-Spolaor\altaffilmark{1,2},
S.~Chatterjee\altaffilmark{5},
J.~M.~Cordes\altaffilmark{5},
N.~J.~Cornish\altaffilmark{7},
F.~Crawford\altaffilmark{8},
H.~T.~Cromartie\altaffilmark{9},
K.~Crowter\altaffilmark{10},
M.~DeCesar\altaffilmark{$\dagger$11},
P.~B.~Demorest\altaffilmark{12},
T.~Dolch\altaffilmark{13},
J.~A.~Ellis\altaffilmark{$\dagger$14,1,2},
R.~D.~Ferdman\altaffilmark{15},
E.~C.~Ferrara\altaffilmark{16},
E.~Fonseca\altaffilmark{17},
N.~Garver-Daniels\altaffilmark{1,2},
P.~Gentile\altaffilmark{1,2},
D.~Good\altaffilmark{10},
J.~S.~Hazboun\altaffilmark{$\dagger$18},
A.~M.~Holgado\altaffilmark{19},
E.~A.~Huerta\altaffilmark{19},
K.~Islo\altaffilmark{20},
R.~Jennings\altaffilmark{5},
G.~Jones\altaffilmark{21},
M.~L.~Jones\altaffilmark{20},
D.~L.~Kaplan\altaffilmark{20},
L.~Z.~Kelley\altaffilmark{22},
J.~S.~Key\altaffilmark{18},
M.~T.~Lam\altaffilmark{23,24,1,2},
T.~J.~W.~Lazio\altaffilmark{25,26},
L.~Levin\altaffilmark{27},
D.~R.~Lorimer\altaffilmark{1,2},
J.~Luo\altaffilmark{17},
R.~S.~Lynch\altaffilmark{28},
D.~R.~Madison\altaffilmark{$\dagger$1,2},
M.~A.~McLaughlin\altaffilmark{1,2},
S.~T.~McWilliams\altaffilmark{1,2},
C.~M.~F.~Mingarelli\altaffilmark{29,30},
C.~Ng\altaffilmark{31},
D.~J.~Nice\altaffilmark{11},
T.~T.~Pennucci\altaffilmark{$\dagger$32},
N.~S.~Pol\altaffilmark{1,2},
S.~M.~Ransom\altaffilmark{9,33},
P.~S.~Ray\altaffilmark{34},
X.~Siemens\altaffilmark{35,20},
J.~Simon\altaffilmark{25,26},
R.~Spiewak\altaffilmark{20,36},
I.~H.~Stairs\altaffilmark{10},
D.~R.~Stinebring\altaffilmark{37},
K.~Stovall\altaffilmark{$\dagger$12},
J.~K.~Swiggum\altaffilmark{$\dagger$20},
S.~R.~Taylor\altaffilmark{38,25,26},
M.~Vallisneri\altaffilmark{25,26},
R.~van~Haasteren\altaffilmark{$\ddagger$26},
S.~J.~Vigeland\altaffilmark{20},
C.~A.~Witt\altaffilmark{1,2},
W.~W.~Zhu\altaffilmark{39}\\
(The NANOGrav Collaboration)\altaffilmark{$\star$}}

\affil{$\star$Author order alphabetical by surname}
\affil{$^{1}$Department of Physics and Astronomy, West Virginia University, P.O.~Box 6315, Morgantown, WV 26506, USA}
\affil{$^{2}$Center for Gravitational Waves and Cosmology, West Virginia University, Chestnut Ridge Research Building, Morgantown, WV 26505, USA}
\affil{$^{3}$X-Ray Astrophysics Laboratory, NASA Goddard Space Flight Center, Code 662, Greenbelt, MD 20771, USA}
\affil{$^{4}$Department of Physics and Astronomy, Widener University, One University Place, Chester, PA 19013, USA}
\affil{$^{5}$Department of Astronomy, Cornell University, Ithaca, NY 14853, USA}
\affil{$^{6}$Cornell Center for Advanced Computing, Ithaca, NY 14853, USA}
\affil{$^{7}$Department of Physics, Montana State University, Bozeman, MT 59717, USA}
\affil{$^{8}$Department of Physics and Astronomy, Franklin \& Marshall College, P.O.~Box 3003, Lancaster, PA 17604, USA}
\affil{$^{9}$University of Virginia, Department of Astronomy, P.O.~Box 400325, Charlottesville, VA 22904, USA}
\affil{$^{10}$Department of Physics and Astronomy, University of British Columbia, 6224 Agricultural Road, Vancouver, BC V6T 1Z1, Canada}
\affil{$^{11}$Department of Physics, Lafayette College, Easton, PA 18042, USA}
\affil{$^{12}$National Radio Astronomy Observatory, 1003 Lopezville Rd., Socorro, NM 87801, USA}
\affil{$^{13}$Department of Physics, Hillsdale College, 33 E.~College Street, Hillsdale, Michigan 49242, USA}
\affil{$^{14}$Infinia ML, 202 Rigsbee Avenue, Durham, NC 27701, USA}
\affil{$^{15}$Department of Physics, University of East Anglia, Norwich, UK}
\affil{$^{16}$NASA Goddard Space Flight Center, Greenbelt, MD 20771, USA}
\affil{$^{17}$Department of Physics, McGill University, 3600  University St., Montreal, QC H3A 2T8, Canada}
\affil{$^{18}$University of Washington Bothell, 18115 Campus Way NE, Bothell, WA 98011, USA}
\affil{$^{19}$NCSA and Department of Astronomy, University of Illinois at Urbana-Champaign, Urbana, Illinois 61801, USA}
\affil{$^{20}$Center for Gravitation, Cosmology and Astrophysics, Department of Physics, University of Wisconsin-Milwaukee,\\ P.O.~Box 413, Milwaukee, WI 53201, USA}
\affil{$^{21}$Department of Physics, Columbia University, New York, NY 10027, USA}
\affil{$^{22}$Center for Interdisciplinary Exploration and Research in Astrophysics (CIERA), Northwestern University, Evanston, IL 60208}
\affil{$^{23}$School of Physics and Astronomy, Rochester Institute of Technology, Rochester, NY 14623, USA}
\affil{$^{24}$Laboratory for Multiwavelength Astronomy, Rochester Institute of Technology, Rochester, NY 14623, USA}
\affil{$^{25}$Jet Propulsion Laboratory, California Institute of Technology, 4800 Oak Grove Drive, Pasadena, CA 91109, USA}
\affil{$^{26}$Theoretical AstroPhysics Including Relativity (TAPIR), MC 350-17, California Institute of Technology, Pasadena, California 91125, USA}
\affil{$^{27}$Jodrell Bank Centre for Astrophysics, University of Manchester, Manchester, M13 9PL, United Kingdom}
\affil{$^{28}$Green Bank Observatory, P.O.~Box 2, Green Bank, WV 24944, USA}
\affil{$^{29}$Department of Physics, University of Connecticut, 196 Auditorium Road, U-3046, Storrs, CT 06269-3046, USA}
\affil{$^{30}$Center for Computational Astrophysics, Flatiron Institute, 162 Fifth Avenue, New York, NY 10010, USA}
\affil{$^{31}$Dunlap Institute for Astronomy and Astrophysics, University of Toronto, 50 St. George St., Toronto, ON M5S 3H4, Canada}
\affil{$^{32}$Hungarian Academy of Sciences MTA-ELTE ``Extragalatic Astrophysics Research Group'', Institute of Physics, E\"{o}tv\"{o}s Lor\'{a}nd University, P\'{a}zm\'{a}ny P. s. 1/A, 1117 Budapest, Hungary}
\affil{$^{33}$National Radio Astronomy Observatory, 520 Edgemont Road, Charlottesville, VA 22903, USA}
\affil{$^{34}$Naval Research Laboratory, Washington DC 20375, USA}
\affil{$^{35}$Department of Physics, Oregon State University, Corvallis, OR 97331, USA}
\affil{$^{36}$Centre for Astrophysics and Supercomputing, Swinburne University of Technology, PO Box 218, Hawthorn, VIC 3122, Australia}
\affil{$^{37}$Department of Physics and Astronomy, Oberlin College, Oberlin, OH 44074, USA}
\affil{$^{38}$Department of Physics and Astronomy, Vanderbilt University, 2301 Vanderbilt Place, Nashville, TN 37235, USA}
\affil{$^{39}$CAS Key Laboratory of FAST, Chinese Academy of Science, Beijing 100101, China}

\affil{$^{\dagger}$ NANOGrav Physics Frontiers Center Postdoctoral Fellow}
\affil{$^{\ddagger}$ currently employed at Microsoft Corporation}
\email[${\color{magenta}\S}$ Corresponding author email: ]{paul.baker@nanograv.org}

\begin{abstract}
The mergers of supermassive black hole binaries (SMBHB) promise to be incredible sources of gravitational waves (GW).
While the oscillatory part of the merger gravitational waveform will be outside the frequency sensitivity range of pulsar timing arrays (PTA),
the non-oscillatory GW memory effect is detectable.
Further, any burst of gravitational waves will produce GW memory, making memory a useful probe of unmodeled exotic sources and new physics.
We searched the North American Nanohertz Observatory for Gravitational Waves (NANOGrav) 11-year data set for GW memory.
This dataset is sensitive to very low frequency GWs of $\sim3$ to $400$ nHz (periods of $\sim11$ yr $-$ $1$ mon).
Finding no evidence for GWs, we placed limits on the strain amplitude of GW memory events during the observation period.
We then used the strain upper limits to place limits on the rate of GW memory causing events.
At a strain of $2.5\times10^{-14}$, corresponding to the median upper limit as a function of source sky position,
we set a limit on the rate of GW memory events at $<0.4$ yr$^{-1}$.
That strain corresponds to a SMBHB merger with reduced mass of $\eta M \sim 2\times10^{10}\Msun$ and inclination of $\iota=\pi/3$ at a distance of 1 Gpc.
\par
As a test of our analysis, we analyzed the NANOGrav 9-year data set as well.
This analysis found an anomolous signal, which does not appear in the 11-year data set.
This signal is not a GW, and its origin remains unknown.
\end{abstract}

\section{Introduction}
\label{sec:intro}
%!TEX root = nanograv_11yr_bwm.tex

% GW memory
Non-oscillatory gravitational wave (GW) effects have been known since the 1970s \citep{zp1974, bg1985, bt1987}.
For cases like supernovae explosions \citep{bh1996} or hyperbolic passages of massive bodies \citep{tw1978},
the non-oscillatory motion of the sources on unbound trajectories is encoded in the GWs as linear GW memory.
For systems with purely oscillatory source motion, like binary black hole (BBH) inspirals,
the GWs themselves follow unbound trajectories, generating non-linear GW memory \citep{c1991, ww1991, bd1992, t1992}.
\par

GW memory is a permanent change to the spacetime metric, contributing a DC component to the GW waveform.
Non-linear memory builds throughout the whole history of a system's evolution,
with the largest accumulation of memory occurring during periods of maximal GW emission.
For compact binary sources, a burst of GW memory is produced during the highly relativistic merger.
While non-linear memory is sourced at the 2.5 post-Newtonian order,
it enters the GW amplitude evolution at the leading, Newtonian order \citep{abiq2004}, making the memory effect tantalizingly detectable.
While the GW memory accumulated during a binary inspiral can be calculated using the post-Newtonian formalism \citep{ww1991,abiq2004},
it was not until \citet{f2009a, f2009b, f2011} estimated the memory effect all the way through binary merger using an effective-one-body approach that interest in the subject was revitalized.
\par

% PTAs
Pulsars act as highly stable galactic clocks \citep[and references therein]{l2008}.
Their stability allows one to detect small changes in the arrival times of pulses caused by the passage of GWs between the pulsar and observer \citep{s1978, d1979}.
Long term pulsar timing campaigns provide sensitivity to low frequency gravitational waves with periods of months to years.
A pulsar timing array (PTA) combines observational data from multiple pulsars boosting sensitivity to common effects like GWs \citep{fb1990, l2015}.
Supermassive black hole binaries (SMBHB) are the most promising sources of GWs for PTAs.
Inspiraling SMBHB could be detected as individual, resolvable sources or as a stochastic background of many overlapping sources \citep{h1994, jb2003, svv2009}.
\par

While the inspiral phase of a SMBHB emits GWs detectable by PTAs, the final merger phase emits GWs that are too high frequency for PTAs to detect
(order days$^{-1}$ for $\sim10^{9} \Msun$ systems).
Despite this, the non-linear GW memory associated with the merger could potentially be resolved independently of the oscillatory component \citep{s2009, vHl2010, pbp2010, cj2012, mcc2014}.
Further, every GW producing system will produce non-linear memory,
so searches for GW memory could uncover exotic sources of GWs or even new physics.
Past studies by the North American Nanohertz Observatory for Gravitational Waves \citep[NANOGrav,][]{ng.2013} and the Parkes Pulsar Timing Array \citep[PPTA,][]{ppta.2013} have searched for and placed limits on GW memory in \citet[][hereafter \citetalias{ng5.bwm}]{ng5.bwm} and \citet{whc+2015}.
Additionally, \citet{mzh+2016} used PPTA data to constrain GW memory signals originating in five nearby galaxy clusters.
\par

As a DC effect, detection prospects for memory have been considered for experiments spanning the GW spectrum.
\citet{f2009a, f2009b, f2011} discussed detection prospects for both LIGO \citep{ligo2015} and LISA \citep{lisa2017}.
As mentioned above, the PTA community quickly realized the memory detection potential of their low-frequency GW experiments.
More recently, \citet{ltl+2016}, \citet{mtl2017}, and \citet{ttl+2018} have considered an approach to detecting GW memory with LIGO by stacking data from several BBH detections.
\citet{mcc2017} have considered the detection prospects of GW memory originating in globular clusters.
\citet{cbv+2014} point out that GW memory accompanying bursts of GWs at high redshift could be detectible.
The ubiquitous nature of GW memory production makes it an excellent discovery tool capable of probing new and exotic physics.
\par

NANOGrav recently published its 11-year data release \citep[][hereafter \citetalias{ng11}]{ng11}.
Using this data set the NANOGrav collaboration has placed limits on a stochastic background of GWs \citep[][hereafter \citetalias{ng11.gwb}]{ng11.gwb}
and on continuous GWs from individual inspiraling SMBH binaries \citep[][hereafter \citetalias{ng11.cw}]{ng11.cw}.
In this work, we search the NANOGrav 9-year \citep[][hereafter \citetalias{ng9}]{ng9} and 11-year (\citetalias{ng11}) data sets for GW memory.
Our primary reported results use \citetalias{ng11}.
\par

\section{The NANOGrav $9$-year and $11$-year Data Sets}
\label{sec:data}
%!TEX root = nanograv_11yr_bwm.tex

For this analysis we used the NANOGrav 9-year (\citetalias{ng9}) and 11-year data sets (\citetalias{ng11}).
The NANOGrav 9-year data set contains the times of arrival (TOAs) for 37 pulsars observed between 2004 and 2013.
The NANOGrav 11-year data set extends the 9-year data set, including TOAs for 45 pulsars with observations extending to 2015.
Several pulsars have been added to the array since regular observations began.
Despite their names, not all pulsars in the 9-year and 11-year datasets have observations covering the whole timespan.
For our analysis of the 11-year data set we used only the 34 pulsars with a minimum of 3 years of observations.
For more details on NANGrav observations and data reduction see \citetalias{ng9} and \citetalias{ng11}.
\par

NANOGrav observations use two radio telescopes:
the 100-m Robert C.~Byrd Green Bank Telescope (GBT) of the Green Bank Observatory in Green Bank, West Virginia;
the 305-m William E. Gordon Telescope (Arecibo) of Arecibo Observatory in Arecibo, Puerto Rico.
Prioritizing Arecibo's better sensitivity, all pulsars visible to Arecibo ($0^\circ < \delta < 39^\circ$) were observed with it.
Those outside Arecibo's declination range were observed with GBT.
Two pulsars, PSRs~J1713+0747 and B1937+21, were observed with both.
We observed most pulsars once a month.
In 2013 we began a high-cadence observing campaign aimed to increase our sensitivity to individual SMBH binary sources \citep{blf2011, cal+14}.
Seven pulsars were observed weekly: PSRs~J1713+0747 and J1909$-$3744 with GBT;
PSRs~J0030+0451, J1640+2224, J1713+0747, J2043+1711, and J2317+1439 with Arecibo.
\par

In order to measure pulse dispersion due to the interstellar medium (ISM) we observed each pulsar at multiple radio frequencies.
At GBT each pulsar was observed with both the 820 MHz and 1.4 GHz receivers.
These two observations were typically separated by a few days due to mechanical and scheduling constraints.
At Arecibo each pulsar was observed with the 1.4 GHz receiver
and one of 430 MHz or 2.3 GHz receiver depending on the properties of the individual pulsars.
The Arecibo observations are made back-to-back with the second frequency observation beginning minutes after the first completes.
The telescopes' backend instrumentation systems were upgraded between 2010 and 2012.
Earlier data were recorded using the 64 MHz bandwidth ASP (Arecibo) and GASP (GBT) systems.
Newer data were recorded with the wideband PUPPI (Arecibo) and GUPPI (GBT) systems.
During the transition data were simultaneously recorded with both systems for verification; however, only the data from the newer system is included in the release.
\par

We fit a timing model to each pulsar's observed TOAs using \texttt{tempo}\footnote{\href{http://tempo.sourceforge.net}{tempo.sourceforge.net}} and \texttt{tempo2}\footnote{\href{https://bitbucket.org/psrsoft/tempo2.git}{bitbucket.org/psrsoft/tempo2.git}} \citep{hem06,ehm06}.
The timing models for all pulsars include the spin period, spin period derivative, sky location, distance, and proper motion.
Pulsars in binaries have additional Keplerian and post-Keplerian parameters describing the binary motion.
The timing models also account for dispersion measure variations, as a piece-wise offset from the mean for each observing epoch, as first discussed in \citet[][hereafter \citetalias{ng5}]{ng5}.
\par

\section{Data Analysis Methods}
\label{sec:analysis}
%!TEX root = nanograv_11yr_bwm.tex

We present the first fully Bayesian search for GW memory with a PTA.
This work represents a leap forward in analysis sophistication over NANOGrav's previous search for GW memory \citepalias{ng5.bwm}.

\subsection{Model Overview}
We modeled the residual pulse time of arrival, $\delta \mathbf{t}$, for a particular pulsar as the sum of stochastic and deterministic components
\begin{align}
    \delta\mathbf{t} &= \mathbf{s} + T\mathbf{b} + \mathbf{n}.
    \label{eq:fullmod}
\end{align}
In this framework $\mathbf{s}$ represents deterministic effects such as those from gravitational waves or a solar system ephemeris (SSE) model.
$T\mathbf{b}$ are stochastic processes described by a Gaussian process model:
$T$ is the design matrix of basis functions for the models,
and $\mathbf{b}$ are the model coefficients.
This Gaussian process model was used for low-frequency (red), intrinsic pulsar noise
and to account for uncertainty in the pulsar timing model.
White noise sources are given by $\mathbf{n}$, including template fitting uncertainty and radio frequency correlated pulse jitter noise \citep[see][and references therein]{cs2010}.
\par

\subsection{Gravitational Wave Model}
\label{sec:analysis.gwmod}
Non-linear GW memory is believed to accompany the oscillatory GWs produced by compact binaries.
\citet{f2009a} computed the total accumulated memory during BBH inspiral and merger, finding\footnote{in geometric units where $G=c=1$}
\begin{align}
	\hmem &= \frac{1}{24} \frac{\eta M}{R} \sin^2 \iota \left(17 +\cos^2 \iota \right) \left[ \frac{\Delta E_\mathrm{rad}}{\eta M} \right], \\
	\frac{\Delta E_\mathrm{rad}}{\eta M} &\sim 1 - \sqrt{8}/3 \sim 0.06,
	\label{eq:strain}
\end{align}
where $M=m_1+m_2$ is the binary total mass;
$\eta=m_1 m_2 / M^2$ is the reduced mass ratio;
$R$ is the co-moving distance to the source;
$\iota$ is the binary inclination;
and $\Delta E_\mathrm{rad}$ is the radiated energy, which is approximated in \autoref{eq:strain} following \citet{lczn2010}.
Assuming a modest inclination $\iota = \pi/3$,
\begin{align}
	\hmem \sim 1.5 \times 10^{-15} \, \left(\frac{\eta M}{10^9 \Msun} \right) \left(\frac{\mathrm{Gpc}}{R}\right).
\end{align}
For a non-precessing source, the GW memory signal is purely linearly polarized.
Following the usual coordinate conventions for BBH inspiral and merger, the memory is ``$+$'' polarized in the source frame.
The observed polarization angle will depend on the specific source-detector geometry.
\par

We implemented the same memory model used by \citet{vHl2010}, \citet{pbp2010}, \citet{mcc2014}, \citet{whc+2015}, and \citetalias{ng5.bwm}.
This model treats the GW memory as a step function that turns on (and off) as the wavefront passes by the Earth at time $t_0$ (and the pulsar at time $t_i$).
This step in the spacetime metric causes a change in the distance between the pulsar and the Earth.
Each radio pulse will arrive progressively more late (or early depending on the sign of the step) compared to the expected TOA.
The response to GW memory in the timing residuals will therefore be a linear increase (or decrease).
\par

For the amplitude of the step we use $\hmem$ directly, bypassing the source specific amplitude dependencies ($M$, $\eta$, $R$, $\iota$).
This model ignores the details of memory accumulation.
We assume that the bulk of the GW memory arises from a transient burst of GWs at a timescale shorter than our $\sim$monthly pulsar observations.
While PTA searches for GW memory were originally motivated by BBH systems, this generic model is agnostic to the source.
\par

We choose to write the GW memory model in a form slightly different than previous work to better illuminate our search parameters.
The GW's effect on the time of arrival of pulses from the $i$th pulsar is given by
\begin{align}
    s_i(t) &= \hmem\, B_i(\hat{k},\psi; \hat{n}_i) \times \nonumber \\
     &\, \left[(t-t_0)\,\Theta(t-t_0) - (t-t_i)\,\Theta(t-t_i)\right],
     \label{eq:sigmod}
\end{align}
where $\hmem$ is the GW strain of the memory;
$B_i$ is the angular response of the pulsar, which depends on its sky position $\hat{n}_i$ and the sky position $\hat{k}(\theta,\phi)$ and polarization $(\psi)$ of the source \citep{ew1975};
$t_0$ is the time that the GW wavefront passes the Earth,
$t_i$ is the retarded time for the GW wavefront passing the pulsar (GW passage time corrected for signal travel time from pulsar to Earth),
and $\Theta$ is the Heaviside step function.
This model assumes a plane-fronted GW with $R\gg\ell_i$, i.e., the distance to the source is much greater than the distance to any pulsar.
The sky positions of the pulsars are very well constrained by the timing model, and we took these to be known exactly.
\par

The final factor of \autoref{eq:sigmod} contains the so called \textit{Earth term} and \textit{pulsar term}, describing the state of the GWs at each location.
With the typical pulsar distance being $\sim$kpc the Earth and pulsar terms will not both fall within our 11-year observation window unless the propagation direction of the GW is nearly perpendicular to the Earth-pulsar separation.
The source positions which result in both the Earth and pulsar term being active in the observation time represent a small but non-negligible fraction of the sky \citep{p2012}.
Further, a search including both the Earth and pulsar term would potentially add a small amount of signal-to-noise for these sources (\citealt{cj2012}; \citetalias{ng5.bwm}).
However, including the pulsar term would greatly complicate the analysis owing to the poorly constrained Earth-pulsar distances.
For this reason, we did not do a simultaneous search for both the Earth and pulsar terms.
Instead, we undertook two separate analyses:
\begin{itemize}
    \item a search for the Earth term only
        \begin{align}
            s_i(t) &= \hmem\, B_i(\hat{k},\psi; \hat{n}_i)\, (t-t_0)\,\Theta(t-t_0).
            \label{eq:sigmod.earth}
        \end{align}
        This search combined data from all pulsars and had five free parameters in the GW model $(\hmem, t_0, \theta, \phi, \psi)$.
    \item a search for the pulsar term only
        \begin{align}
            s_i(t) &= s\, \hopt \, (t-t_i)\,\Theta(t-t_i),
            \label{eq:sigmod.snglpsr}
        \end{align}
        where $\hopt$ is the strain amplitude assuming an optimally oriented source
        and $s$ is the sign $(+/-)$ of the memory effect.
        The pulsar term search was conducted on each pulsar individually and had three free parameters in the GW model $(\hopt, t_i, s)$.
        The signal model was modified for the pulsar term search,
        because the angular response $B$ is completely covariant with the GW amplitude $\hmem$.
        When using a single pulsar, the extrinsic parameters of the source $(\hat{k}, \psi)$ are not constrained.
\end{itemize}
\par

It is nearly impossible to make a confident detection with a single pulsar term search,
because the GW memory signal is nearly indistinguishable from an intrinsic pulsar glitch \citep{vHl2010}.
Even though it would be hard to trust a GW detection from a single pulsar search,
we can still use non-detection in the pulsar term search to set upper limits.
The individual pulsar term searches cover many independent time periods, making pulsar term limits especially useful for constraining the rate of GW memory producing events.
\par

There are a small fraction of source locations that would result in the GW passing multiple pulsars but not the Earth during our observation.
Analyzing multiple pulsar terms without the Earth term to cover these cases, could result in a small boost to signal-to-noise.
As in the Earth term search, the poorly constrained Earth-pulsar distances would greatly complicate any multiple pulsar term search.
For this reason, we performed only single pulsar term searches.
We did not search over cases with multiple simultaneous pulsar terms for simplicity.
\par

\subsection{White Noise Model}
We used the pulsar noise model described in \citet{lcc+2016} and also used in \citetalias{ng11} and \citetalias{ng11.gwb}.
The white noise is parametrized in each pulsar per each observing system, $k$ (each unique combination of telescope frontend and backend hardware, e.g., L-band GUPPI).
This noise $\mathbf{n}$ is defined by a covariance matrix with each TOA specified by its observation time $t$ and radio frequency $\nu$
\begin{align}
    N_{\nu\nu^\prime\,tt^\prime\,k} &= \delta_{tt^\prime} \left[
      \delta_{\nu\nu^\prime} \left(
        {\mathcal{F}_k}^2\sigma^2 + {\mathcal{Q}_k}^2
      \right) + {\mathcal{J}_k}^2
    \right],
    \label{eq:noise-wh}
\end{align}
where $\sigma$ is the pulse template fitting uncertainty;
$\mathcal{F}_k$ is `EFAC', an additional scaling factor (in practice $\sim$1 for all);
$\mathcal{Q}_k$ is `EQUAD', an additional variance added in quadrature, `EQUAD';
$\mathcal{J}_k$ is `ECORR', a component that is correlated between different radio frequency channels for a given observation, but not correlated from one observation to the next.
$\mathcal{J}$ includes pulse jitter noise.
Finally, $\delta$ is the Kronecker delta.
The parameters $\mathcal{F}$, $\mathcal{Q}$, and $\mathcal{J}$ account for additional noise which is empirically observed, but not accounted by pulse template fitting uncertainty $\sigma$ alone \citep{lcc+2016}.
The `E' names refer to the parameter names given in the \texttt{tempo} and \texttt{tempo2} pulsar timing software.
\par

The white noise covariance matrix $N$ is block diagonal.
The $\sigma$s, $\mathcal{F}$s, and $\mathcal{Q}$s run down the diagonal,
and the $\mathcal{J}$s form blocks connecting each frequency channel from the same observation.

\subsection{Gaussian Process Models}
We modeled the remaining stochastic processes as Gaussian processes
(see Appendix C of \citetalias{ng9} and section 3 of \citealt{ng9.gwb} for more discussion of this methodology).
In this framework each process is defined by $T$, a design matrix of basis functions,
$\mathbf{b}$, a vector of basis coefficients,
and $B$, a covariance matrix defining the Gaussian priors on $\mathbf{b}$.
\par

We defined the intrinsic pulsar red noise as a Fourier-basis Gaussian process.
The design matrix $T_\mathrm{red}$ contains the sine and cosine basis functions,
and the coefficients $\mathbf{b}_\mathrm{red} = (a, b)_j$ are the Fourier coefficients such that
\begin{align}
    T_\mathrm{red}\mathbf{b}_\mathrm{red} &= \sum_{j=1}^N \left[
        a_j \sin\left( 2\pi\, f_j t \right) +
        b_j \cos\left( 2\pi\, f_j t \right)
    \right]
\end{align}
We restricted the sum to the first $N=30$ Fourier components (starting at the inverse observation time),
as this model is for \textit{low-frequency} noise unaccounted elsewhere.
\par
The red noise is modeled as a power-law spectrum
\begin{align}
    P(f_j) &= {A}^2 \left( \frac{f_j}{\fyr} \right)^{-\gamma} \, \mathrm{yr}^3,
\end{align}
where $A$ is the characteristic amplitude at the reference frequency of $\fyr$,
and $\gamma$ is the spectral slope.
The spectral shape defines the priors on the Fourier coefficients.
The individual Fourier components are approximated as uncorrelated, so the resulting covariance is diagonal:
\begin{align}
    (B_\mathrm{red})_{ii} &= P(f_i).
\end{align}
\par

We also defined the timing model uncertainty in the Gaussian process framework.
The design matrix, $T_\mathrm{tm}$, has columns that are the timing model linearized around the best fit parameters,
and the basis coefficients, $\mathbf{b}_\mathrm{tm}$, are small offsets from the best fit timing parameters.
We placed an unconstrained, uniform prior on the timing offsets by setting the covariance to a diagonal matrix of infinities, $(B_\mathrm{tm})_{ii} = \infty$.
\par

We then concatenated the various Gaussian processes into the form of \autoref{eq:fullmod}.
\begin{align}
    T &=
    \begin{bmatrix}
        T_\mathrm{tm} & T_\mathrm{red}
    \end{bmatrix}, &
    \mathbf{b} &=
    \begin{bmatrix}
        \mathbf{b}_\mathrm{tm}\\
        \mathbf{b}_\mathrm{red}
    \end{bmatrix}, &
     B &=
     \begin{bmatrix}
        B_\mathrm{tm} & \\
         & B_\mathrm{red} \\
     \end{bmatrix}.
\end{align}
Additional Gaussian processes can be cleanly added in the same way.
For instance, one could include a stochastic GW background as a correlated noise source in all pulsars by adding an additional Gaussian process model.
We do not include it in this work,
but this data anlysis framework allows for a stochastic GW background to be added in a straightforward manner in the future.

\subsection{Solar System Ephemeris Model}
\citetalias{ng11.gwb} showed that the NANOGrav 11-year dataset is sensitive to uncertainty in the solar system ephemeris (SSE).
SSE errors can appear as a spatially correlated stochastic process.
As the GW background is also a spatially correlated stochastic effect, there is a natural covariance between the two \citep{thk+2015}.
Because GW memory appears as a transient, deterministic effect, there is little confusion between SSE errors and GW memory.
\par

Despite this, we chose to use the same \bayesephem~model described in \citetalias{ng11.gwb} to mitigate SSE uncertainty in this analysis.
\bayesephem~implements perturbations to a given SSE by varying 11 parameters: the masses of the gas giants (4), the rotation rate about the ecliptic pole (1), and Juptier's orbital elements (6).
We repeated most analyses using two recent JPL SSEs: DE430 and DE436 \citep{DE430, DE436}.
For each SSE we conducted two analyses: holding the SSE fixed and using the \bayesephem~model.
Results reported as ``\bayesephem'' used DE436 as the input SSE before perturbations.
There was no measurable difference between using DE430 and DE436 as the \bayesephem~input.
\par

\subsection{Bayesian analysis}
\subsubsection{Likelihood}
We constructed a Gaussian likelihood based on the white noise covariance.
The model residuals, $\mathbf{r}$, should follow the same distribution as the white noise, $\mathbf{n}$:
\begin{align}
    \mathbf{r} &= \delta\mathbf{t} - T\mathbf{b} - \mathbf{s} \nonumber \\
    p\left(\delta\mathbf{t} \mid \mathbf{b}, \lambda \right) &=
        \frac{ \exp(-\frac{1}{2} \mathbf{r}\cdot N^{-1}\cdot \mathbf{r}) }{\sqrt{\det(2\pi N)}},
\end{align}
where $\lambda$ are the model parameters (red and white noise, deterministic signals),
and $N$ is defined by \autoref{eq:noise-wh}.

Following the scheme of \citet{lah+2013} and \citet{vHv2015}, we can analytically marginalize over the Gaussian process coefficients $\mathbf{b}$,
leaving us with
\begin{align}
    \mathbf{q} &= \delta\mathbf{t} - \mathbf{s} \nonumber \\
    C &= N + TBT^T \nonumber \\
    p\left(\delta\mathbf{t} \mid \lambda \right) &=
        \frac{ \exp(-\frac{1}{2} \mathbf{q}\cdot C^{-1}\cdot \mathbf{q}) }{\sqrt{\det(2\pi C)}}.
    \label{eq:like}
\end{align}
The white $(\mathcal{F}, \mathcal{Q}, \mathcal{J})$ and red $(A, \gamma)$ per-pulsar noise parameters contribute to the final covariance matrix, appearing in $N$ and $B$, respectively.
The \citet{w50} matrix identity can be used to evaluate \autoref{eq:like} efficiently.
Sparse matrix algebra can provide an additional speedup.

The noise parameters that appear in $C$ of \autoref{eq:like} were first fit with individual noise analyses for each pulsar.
For computational efficiency in the GW analyses, we held the white noise parameters fixed to their median values.
The per-pulsar red noise parameters were simultaneously searched with the global GW parameters and SSE parameters, owing to their covariance.
With 34 pulsars the search space for the 11-year Earth term analysis contained 84 dimensions, $(34\times 2_\mathrm{RN}) + 5_\mathrm{GW} + 11_\mathrm{SSE}$.

\subsubsection{Priors}
We prefered ignorance priors for our model parameters, implementing uniform or log-uniform priors for all.
We used the same priors for noise parameters as \citetalias{ng11.gwb} and \citetalias{ng11.cw}.
\par

For detection analysis we follow the philosophy of \citetalias{ng11.gwb}, setting a log-uniform prior on $\hmem$.
This prior is improper for upper limit analysis:
in order to set upper limits on the amplitude of GW memory we must integrate our posterior from a lower bound of $\hmem = 0$.
Following the detection analysis, we reran our pipeline using a uniform prior on $\hmem$ for the purpose of setting upper limits.

\subsubsection{Inference}
We used the Bayes factor for the GW model compared to a noise only model, $\mathcal{B}_\mathrm{gw}$, as our detection statistic.
We calculated Bayes factors using the Savage-Dickey approximation \citep{d71},
\begin{align}
    \mathcal{B}_\mathrm{gw} =
    \frac{\mathcal{E}_\mathrm{gw}}{\mathcal{E}_\mathrm{noise}} \approx
    \lim_{\hmem\rightarrow0}\frac{p\left(\hmem\right)}{p\left(\hmem \mid \delta\mathbf{t}\right)},
\end{align}
which approximates the evidence ratio ($\mathcal{E}_\mathrm{gw} / \mathcal{E}_\mathrm{noise}$) for the GW and noise only models as the ratio of the prior to posterior probability in the limit that GW amplitude goes to zero.
This is calculation uses posterior samples near the low amplitude prior boundary and is much more computationally efficient than a full evidence integral.
The GW model is favored when there is low posterior probability for small $\hmem$, i.e., the posterior for $\hmem$ is peaked away from zero.

\subsection{Software}
Our analysis, like \citetalias{ng11.gwb} and \citetalias{ng11.cw}, used NANOGrav's core data analysis software, \texttt{enterprise}\footnote{\href{https://github.com/nanograv/enterprise}{github.com/nanograv/enterprise}} \citep{enterprise} to compute the posterior probability for our models.
Likelihood evaluations were sped up using sparse matrix algebra
with the \texttt{scikit-sparse}\footnote{\href{https://github.com/scikit-sparse/scikit-sparse}{github.com/scikit-sparse/scikit-sparse}} Python package
and the \texttt{SuiteSparse}\footnote{\href{http://faculty.cse.tamu.edu/davis/suitesparse.html}{faculty.cse.tamu.edu/davis/suitesparse.html}} C library \citep{cholmod}.
We sampled our posterior distribution for Bayesian inference with \texttt{PTMCMCSampler}\footnote{\href{https://github.com/nanograv/enterprise}{github.com/jellis18/PTMCMCSampler}} \citep{ptmcmc}.
We used \texttt{healpy}\footnote{\href{https://github.com/healpy/healpy}{https://github.com/healpy/healpy}} \citep{healpy}
and \texttt{HEALPix}\footnote{\href{https://healpix.jpl.nasa.gov/}{https://healpix.jpl.nasa.gov/}} \citep{healpix} to grid the sky for some analyses.
\par

\section{Results}
\label{sec:results}
%!TEX root = nanograv_11yr_bwm.tex

\subsection{Detection Statistics}
\label{sec:results.detect}
We find no evidence for GW memory in the NANOGrav 11-year dataset.
For the Earth term search we analyzed the 34 pulsars with a minimum observation baseline of three years
(we did not analyize the 11 additional pulsars which had been observed for less than this).
We set a log-uniform prior on GW memory strain amplitude and uniform priors on the other search parameters.
We searched over burst epochs between 2005.7 and 2015.4.
We did not search for GWs in the first 10\% and last 5\% of the dataset,
owing to biases brought about by the loss of sensitivity near the edges of the dataset
(this effect is clearly seen in our amplitude upper limits, e.g.,~\autoref{fig:ULvt.11}).
The Earth term search results in a Bayes factor of $\mathcal{B}_\mathrm{gw}\approx0.7$ for the GW memory model compared to the noise only model.
\par

\begin{figure}
  \includegraphics[width=\columnwidth]{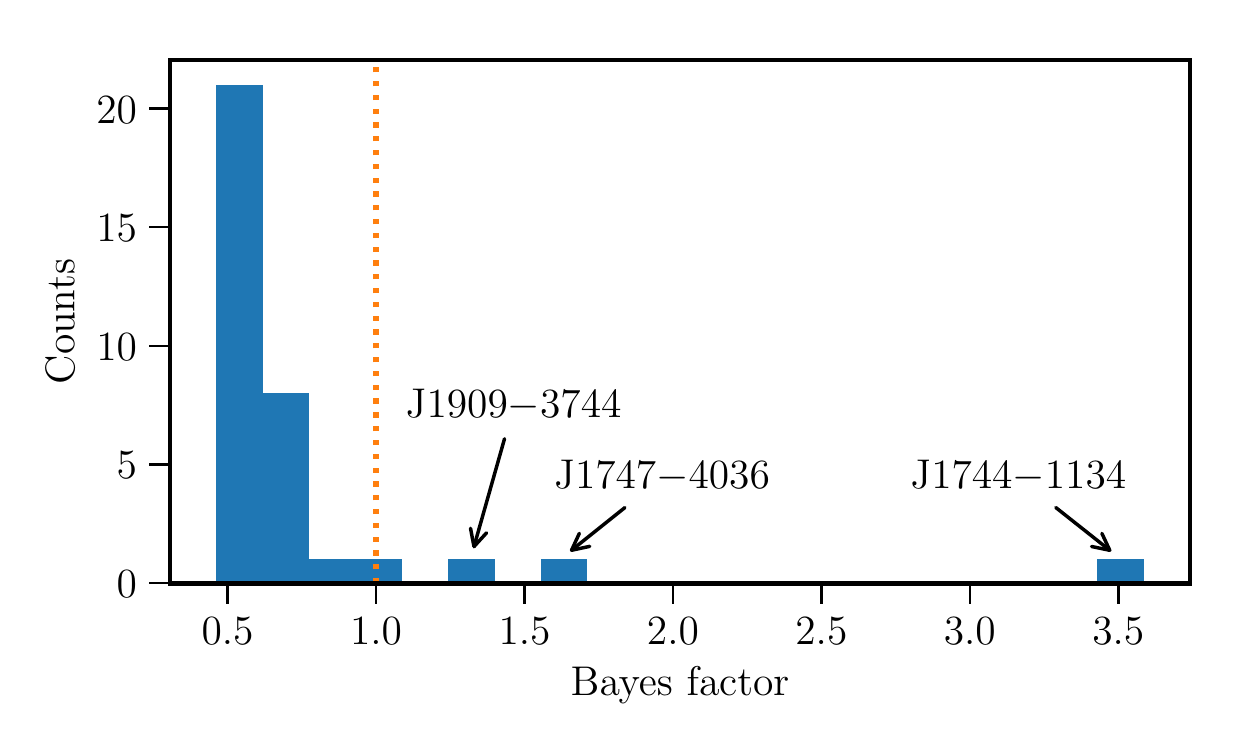}
  \caption{Bayes factor distribution for single pulsar term searches for GW memory.
  Bayes factors less than 1 imply the model without a GW memory burst is preferred.
  Three of the 34 pulsars have a Bayes factor slightly larger than one, meaning the GW memory model is slightly preferred over the noise model.
  None are significant.
  }
  \label{fig:psrBF}
\end{figure}

In the pulsar term search we analyzed data from the same 34 pulsars, individually, calculating the Bayes factor for each.
The results are shown in \autoref{fig:psrBF}.
Most pulsars prefer the model without a GW memory burst, having $\mathcal{B}_\mathrm{gw}<1$.
PSR \psrBF~is the pulsar with the largest Bayes factor, $\mathcal{B}_\mathrm{gw}\approx3.5$.
This Bayes factor is on the threshold of worth mentioning according to the \citet{j1961} scale.
The preferred burst time, $t_0$, occurs near the beginning of \psrBF's observation period, when data were of lower quality (sparser sampling, narrow radio band).
\par

We also searched the NANOGrav 9-year dataset, using the same methods as the 11-year.
In this case we found an anomalous GW memory-like signal at MJD $55422\pm46$, about 2010.6.
If this were truly a GW signal, it would have appeared in the 11-year data analysis, as well.
Seeing as it did not, we can confidently say it must not be a GW memory burst.
Additionally, nearly any modification to the individual pulsar noise modeling reduces the significance of this event considerably.
For further discussion of this anomalous event see \autoref{sec:app.anom}.

\subsection{Upper limits}

Finding no evidence for GW memory in our data we place upper limits on the strain of GW memory events during our observations.
To compute strain upper limits we sampled the log-amplitude of GW strain, placing an exponential prior on this parameter.
This is equivalent to sampling strain amplitude directly with a uniform prior.
Our prior choice biases the posterior toward higher amplitude.
This well known effect results in conservative limits and is not usually a problem.
In our case the non-uniform sensitivity of our PTA combines with the amplitude prior in such a way that the most insensitive times and sky positions dominate the posterior.
\par

For GW amplitudes below the sensitivity of our PTA, the likelihood is flat: changes in signal parameters do not change the likelihood.
In this case the posterior is dominated by the amplitude prior.
The highest probability regions of the posterior will correspond to insensitive source orientations (sky position, polarization) and times, where the GW amplitude can be made largest without affecting the likelihood.
The end result is a posterior that is peaked at our PTA's blind spots.
If we were to naively perform an all sky, all time search and use the 1D marginal posterior for strain amplitude to compute an upper limit, the limit would be dominated by the most insensitive times and source orientations.
This limit would not be a fair representation of our search.
\par

For physical reasons we do not expect GWs to originate from any particular direction.
The non-uniformity of our posterior distribution in source orientation is caused by the amplitude prior.
We can fix this by implementing a non-uniform prior on source orientation which exactly cancels the bias from the amplitude prior.
This type of prior scheme is sometimes called a Malmquist prior, as it corrects for selection effects as \citet{m1922} did for stellar absolute magnitude.
We describe our method to unbias our rate upper limits below.
\par

\begin{figure}
  \includegraphics[width=\columnwidth]{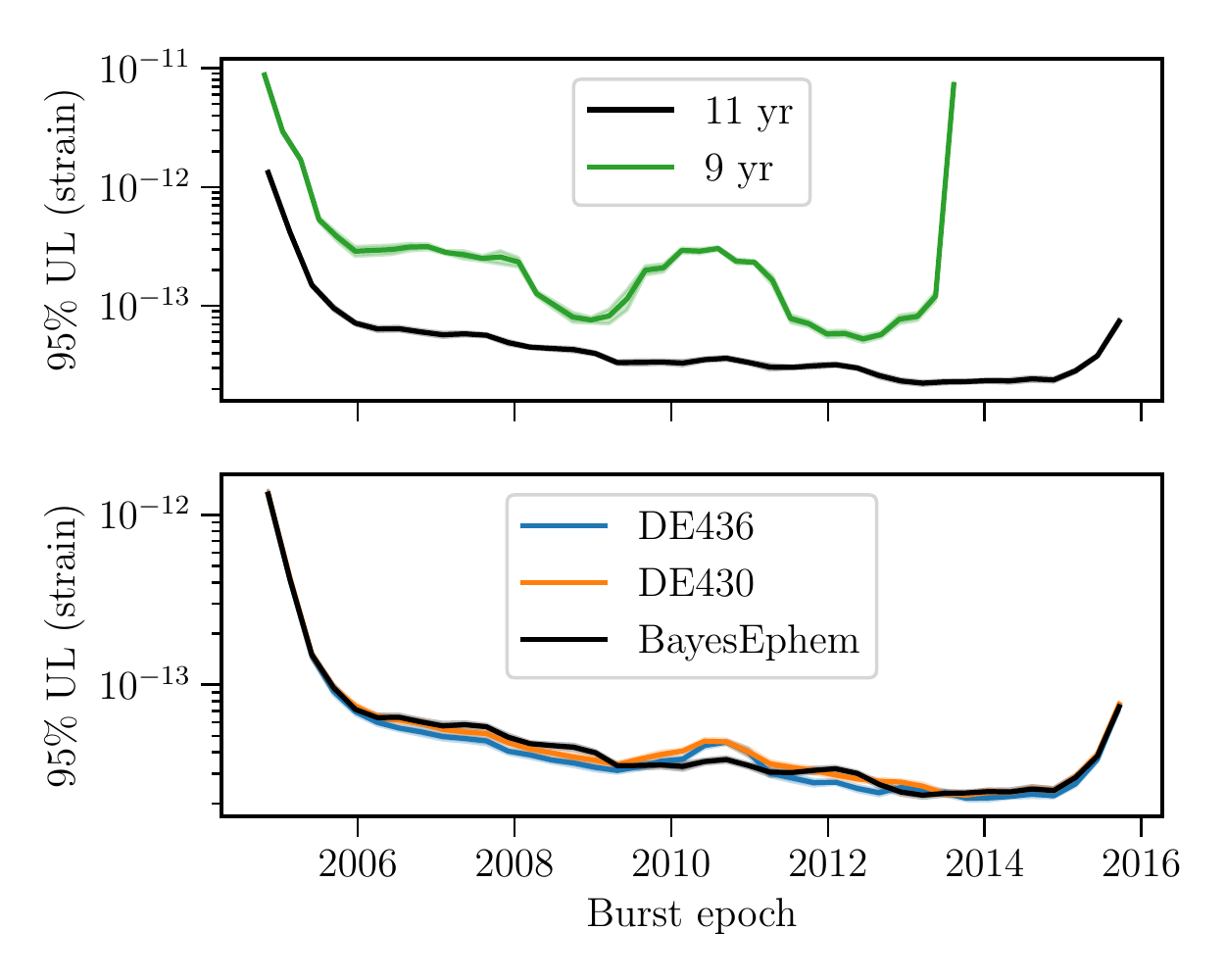}
  \caption{95\% upper limit on gravitational wave memory strain amplitude as a function of burst epoch,
  marginalized over source orientation.
  The very small semi-transparent regions contain the 90\% sampling uncertainty on upper limits.
  \textit{Top}: Comparison of 9-year and 11-year data sets using \bayesephem.
  The elevated upper limit from the 9-year data set during 2010-2012 is a result of the anomolous signal.
  See \autoref{sec:results.detect} and \autoref{sec:app.anom} for further discussion.
  \textit{Bottom}: Comparison of 11-year data set under different SSEs.
  Note that the black curve, 11-year with \bayesephem~is the same in both.
  }
  \label{fig:ULvt.11}
\end{figure}

To place upper limits on the rate of GW memory bursts, we first need strain amplitude upper limits as a function of time.
These are computed by determining a source orientation averaged upper limit for each of 40 time bins.
We drew MCMC samples for each time bin resulting in a posterior biased to insensitive source orientations, as described above.
We then resampled the biased posterior, effectively implementing a \textit{post hoc} prior on source orientation to ensure uniform distributions.
In our resampling scheme, we binned the source sky location using an \texttt{HEALPix} grid with 48 bins, \texttt{nside} $=2$.
We binned the polarization angle into 8 bins from $0$ to $\pi$.
We then drew samples from the biased posterior ensuring an equal number of samples lie in each of these 384 source orientation bins.
By construction the resulting samples are uniformly distributed in source orientation.
The final 1D marginal distribution for strain amplitude is uniformly averaged over source orientation and not biased toward insensitive locations.
The results of this effort are shown in \autoref{fig:ULvt.11}.
We find that our limits are not drastically affected by choice of SSE.
\par

\begin{figure}
  \includegraphics[width=\columnwidth]{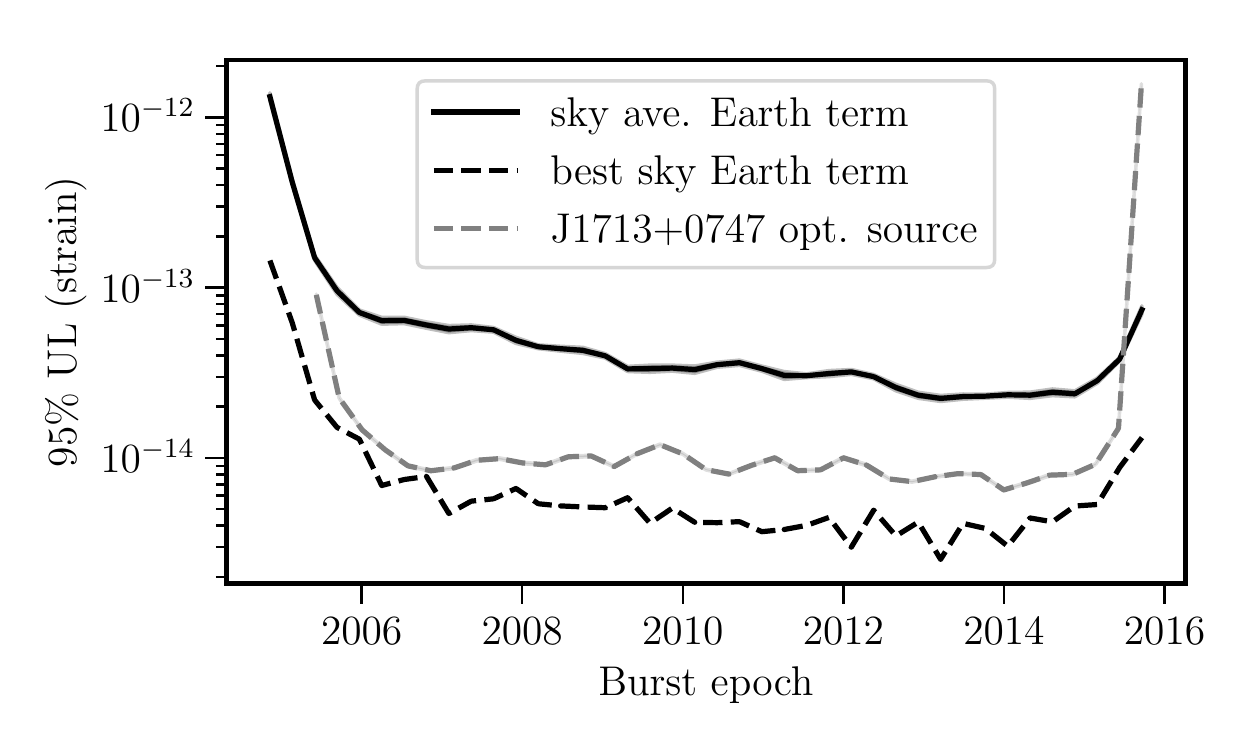}
  \caption{95\% upper limit on gravitational wave memory strain amplitude as a function of burst epoch.
  The three curves show the sky averaged Earth term upper limit (same as \autoref{fig:ULvt.11});
  the Earth term upper limit for the most sensitive sky position for each epoch; and
  the pulsar term upper limit for an optimally oriented source using the most sensitive single pulsar, PSR \psrBestA.
  }
  \label{fig:ULvt.comp}
\end{figure}

To produce a pulsar term limit we conducted the upper limit versus time analysis for each individual pulsar in the array.
First, we computed the upper limit as a function of time for a optimally oriented source.
The upper limit for the most sensitive pulsar, PSR~\psrBestA, is shown in \autoref{fig:ULvt.comp}.
It is compared to the sky averaged Earth term limit, and the limit using the most sensitive sky position from the Earth term search.
We see that for optimally oriented sources the Earth term which combines information from many pulsars is more limiting than the single best pulsar.
\par

An individual pulsar cannot distinguish source orientation as discussed in \autoref{sec:analysis.gwmod}.
In order to set a limit for all sources, we applied a correction factor to account for a source orientation average.
The correction factor comes from analytically marginalizing over source orientation, assuming uniform priors on sky position and polarization angle, and is shown in \autoref{sec:app.ave}.
The choice of prior for the pulsar term sky averaging matches the resampled posterior, ensuring the two can be compared directly.
\par

\begin{figure}
  \includegraphics[width=\columnwidth]{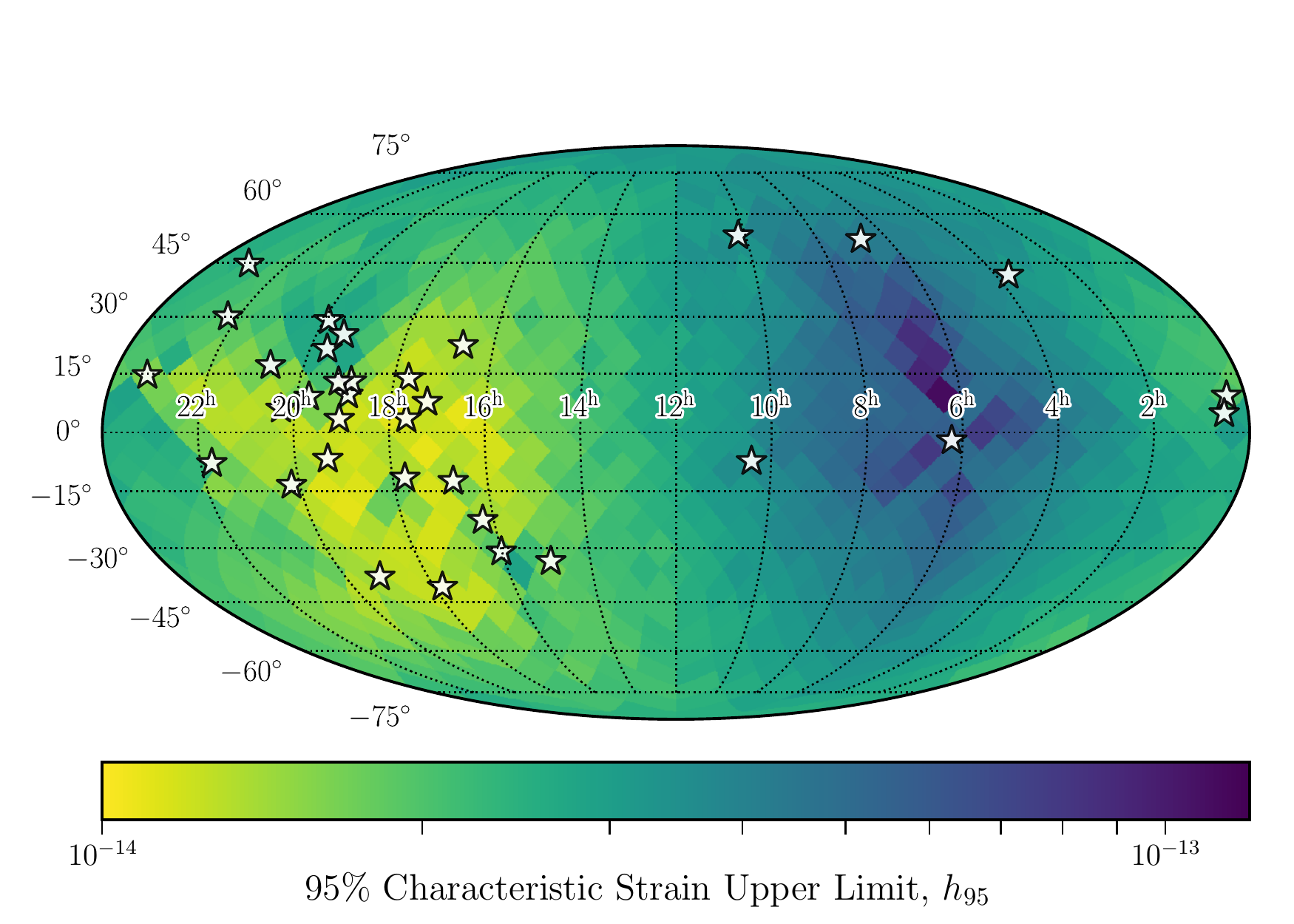}
  \caption{95\% upper limit on gravitational wave memory strain amplitude as a function of sky location of source, using \bayesephem.
  We placed a prior on burst epoch to constrain the analysis to the more recent time span of the data $\sim$2012-2015.
  The low density of pulsars in RA 0-12h makes us much less sensitive to GW memory originating in that hemisphere.
  Stars mark the locations of the 34 pulsars used in this work.
  This map is in equatorial coordinates
  }
  \label{fig:ULvsky.11}
\end{figure}

To explicitly show how the sky position of the source affects the strain upper limit, we conducted a second analysis.
In this case we set a limit for each pixel of an \texttt{nside} $=8$ \texttt{HEALPix} grid (768 sky locations) marginalizing over polarization angle.
Because the sky sensitivity of the PTA changes as new pulsars are added to the array, we focused this analysis to more recent times, $\sim$2012-2015.
We used all of the observed TOAs in this analysis, but only searched for GW memory in that time span.
The results of this analysis are shown in \autoref{fig:ULvsky.11}.
Our PTA is up to an order of magnitude more sensitive to sources originating from the most sensitive sky positions compared to the least.
\par

Note that the optimal source upper limit shown in \autoref{fig:ULvt.comp} dips below $10^{-14}$.
This optimal source includes the optimal polarization angle, while the analysis shown in \autoref{fig:ULvsky.11} marginalizes over polarization angle.
\par

\section{Discussion and Conclusions}
\label{sec:concl}
%!TEX root = nanograv_11yr_bwm.tex

\begin{figure*}
  \includegraphics[width=\textwidth]{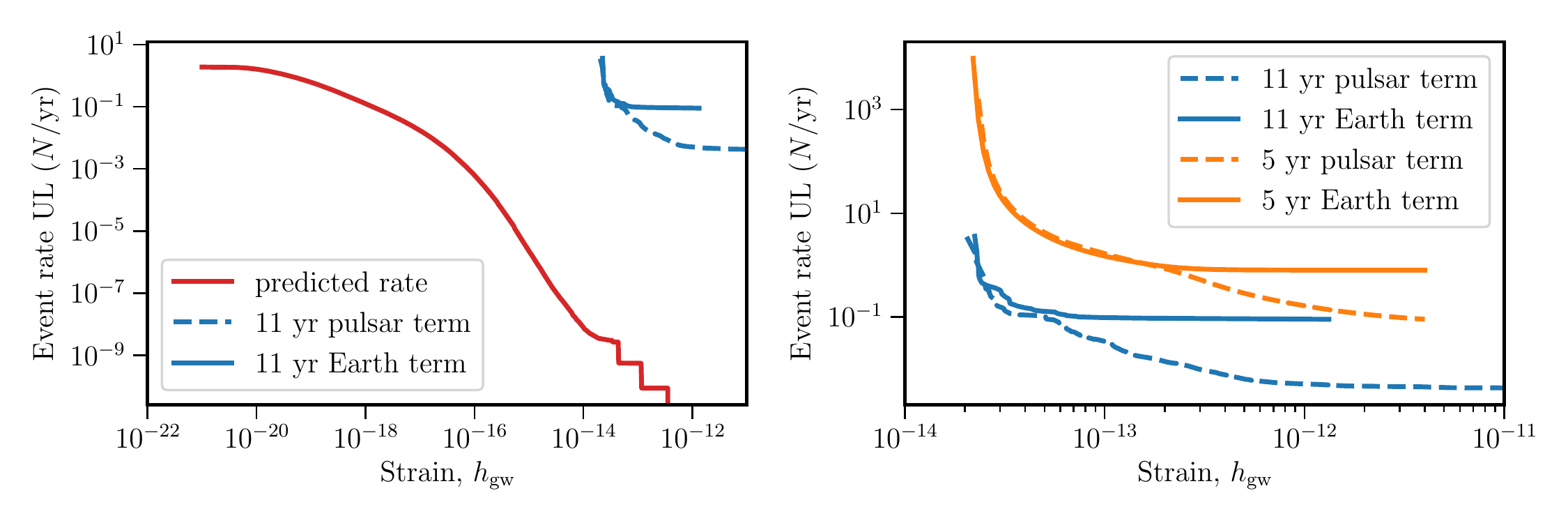}
  \caption{95\% upper limit on the rate of memory causing events as a function of strain amplitude.
  \textit{left}: Comparison of the 11-year data set rate upper limit to the memory rate from SMBHB mergers predicted by \cite{isb+2019}.
  \textit{right}: Comparison of the 11-year data set upper limit to the limits previously published in \citetalias{ng5.bwm}.
  Note that the blue curves are the same in both panels.
  }
  \label{fig:rateUL.comp}
\end{figure*}

From the strain upper limits as a function of burst epoch, we can construct a limit on the rate of GW memory events.
\autoref{fig:rateUL.comp} shows these limits.
The left panel compares the rate limits from this work to those predicted by \citet{isb+2019}, who predicted the rate of memory events from SMBHB mergers for LISA and PTAs.
Based on their analysis using a simulated galaxy stellar mass function from \citet{sbd+2014} and the $M_\mathrm{BH}-M_\mathrm{bulge}$ relation of \citet{mm2013},
we should expect a very small number of SMBHB mergers producing GW memory detectable by PTAs.
The right panel of \autoref{fig:rateUL.comp} compares the limits from this work to the previous published NANOGrav limits of \citetalias{ng5.bwm}.
\par

The median sky position bin from \autoref{fig:ULvsky.11} has an upper limit of $\sim2.5\times10^{-14}$.
Taking this as representative of our Earth term strain sensitivity, this work improves the rate constraints at that strain by more than 2 orders of magnitude, a factor of $\sim160$, relative to \citetalias{ng5.bwm}.
For larger amplitudes, $\gtrsim3\times10^{-13}$ this work improves the \citetalias{ng5.bwm} Earth term limits by a factor of $~10$.
Finally, this work improves the large amplitude pulsar term limit by a factor of $\sim20$ compared to \citetalias{ng5.bwm}.
\par

Our non-detection of GW memory from SMBHB merger should come as no surprise.
While this work was motivated by the prospects of detecting SMBHB merger,
GW memory is a generic feature of all GW producing events.
Our rate limits are presented as agnostic limits on events that produce GW memory of a particular amplitude.
\par

Future searches for GW memory with PTAs are unlikely to detect SMBHB mergers, but other sources can produce GW memory detectable by PTAs.
Because the amplitude of GW memory is proportional to the source mass and inversely proportional to distance, GW memory from much smaller sources could be detectible much closer to home.
For instance, \citet{mcc2017} discussed the prospects of detecting stellar mass compact binary mergers in globular clusters in the Milky Way.
If memory sources are located in the Milky Way, the data analysis methods used will need to be altered.
Nearby sources will violate the plane-wave assumption, so \cite{mcc2017} considered spherically fronted waves.
Depending on the location of the GW memory producing event relative to the pulsars in the array, an event could activate multiple pulsar terms and/or the Earth term.
Searches for these sources will face many of the same challenges that have affected searches for continuous GWs from individual SMBHB sources.
Effictive methods to simultaneously determine the poorly constrained Earth-pulsar separations and the GW parameters is foremost among these problems \citep{cc2010}.
The solution to this problem for GW memory searches will differ from those implemented in continuous GW searches \citep[e.g.,][]{e2013,teg2014} owing to the transient nature of the GW memory signal.
Like in continuous GW searches, incorporating good prior information from pulsar distance measurments will play a crucial role.
Pulsar timing provides distance measurements via parallax for some pulsars (e.g., \citealt{ng9.am}; \citetalias{ng11}).
For pulsars with no timing parallax, an independent distance measurement should be incorporated, if available.
These could come from Very Long Baseline Interferometry \citep[VLBI, e.g.,][]{dtb+2009} or other astrometric experiments \citep[e.g.,][]{mab+2018,jks+2018}.
\par

There remains a rich discovery space for exotic sources of GW memory \citep{cbv+2014}.
As any burst of GWs will produce GW memory, exotic GW producing events such as cosmic strings \citep{dv2001} are possible PTA sources.
Even some non-GW effects, such as a cosmic string crossing between the line of sight from the Earth to a pulsar, exhibit a similar response in pulsar timing data \citep{pt2010}.
Pulsar glitches also produce a signal very similar to GW memory.
While glitches are much more common in canonical pulsars, there are some observations of glitches in millisecond pulsars \citep{cb2004,mjs+2016}.
It is possible to use our limits on GW memory to place limits on glitches in the 34 millisecond pulsars studied.
\par

More generally, searches for transient GWs in PTAs can reveal transient features in the noise that are not currently modeled.
This analysis discovered strange noise features in two of NANOGrav's longest timed pulsars, PSRs \psrAnomalyA~and \psrAnomalyB, in relation to the anomalous signal detected in \citetalias{ng9}.
Both of these pulsars were found to have unmodeled excess noise by \citet{lcc+2017}; however, neither stands out as extraordinary in its noise features in that study from other pulsars in the array.
These newly discovered noise features, along with surprising noise features uncovered in concurent NANOGrav data analysis (\citetalias{ng11.cw}; \citealt{ng11.slice}),
are driving the development of new PTA data analysis techniques.
\par

Looking into the future, searches for GW memory should remain an integral part of the PTA data analysis regime.
These analyses should be implemented on new datasets, like the second data release from the International Pulsar Timing Array \citep[IPTA,][]{ipta.dr2}.
\par

\acknowledgements

\emph{Author contributions.}
% !TEX root = nanograv_11yr_cw.tex
This document is the result of more than a decade of work by the entire NANOGrav collaboration.
We acknowledge specific contributions below.
Z.A., K.C., P.B.D., M.E.D., T.D., J.A.E., R.D.F., E.C.F., E.F., P.A.G., G.J., M.L.J., M.T.L., L.L., D.R.L.,
R.S.L., M.A.M., C.N., D.J.N., T.T.P., S.M.R., P.S.R., R.S., I.H.S., K.S., J.K.S., and W.Z. developed the 11-year data set.
P.T.B. led this analysis and coordinated the paper writing.
J.A.E. and P.T.B. implemented the search algorithms in \texttt{enterprise}.
P.T.B. and K.I. performed the data analysis.
K.A., A.M.H., and N.S.P. conducted preliminary search pipeline testing.
D.R.M., J.A.E., S.R.T., and R.vH. performed an initial analysis of the 9-year data set including an investigation of the anomolous signal.
K.I. and S.B-S. contributed to the astrophysical interpretation.

\emph{Acknowledgments.}
% !TEX root = nanograv_11yr_cw.tex
The NANOGrav project receives support from National Science Foundation (NSF) Physics Frontier Center award \#1430284.
NANOGrav research at UBC is supported by an NSERC Discovery Grant and Discovery Accelerator Supplement and by the Canadian Institute for Advanced Research.
Portions of this research were carried out at the Jet Propulsion Laboratory, California Institute of Technology, under a contract with the National Aeronautics and Space Administration.
P.T.B. acknowledges support from the West Virginia University Center for Gravitational Waves and Cosmology.
M.V. and J.S. acknowledge support from the JPL RTD program.
S.R.T. was partially supported by an appointment to the NASA Postdoctoral Program at the Jet Propulsion Laboratory, administered by Oak Ridge Associated Universities through a contract with NASA.
J.A.E. was partially supported by NASA through Einstein Fellowship grants PF4-150120.
S.B.S. and C.A.W. were supported by NSF award \#1815664.
W.W.Z. is supported by the Chinese Academy of Science Pioneer Hundred Talents Program, the Strategic Priority Research Program of the Chinese Academy of Sciences grant No.~XDB23000000, the National Natural Science Foundation of China grant No.~11690024, and by the Astronomical Big Data Joint Research Center, co-founded by the National Astronomical Observatories, Chinese Academy of Sciences and the Alibaba Cloud.
Portions of this work performed at NRL are supported by the Chief of Naval Research.
The Flatiron Institute is supported by the Simons Foundation.
\par

We are grateful for computational resources provided by the Leonard E Parker Center for Gravitation, Cosmology and Astrophysics at the University of Wisconsin-Milwaukee, which is supported by NSF Grants 0923409 and 1626190.
%This research made use of the Super Computing System (Spruce Knob) at WVU, 
%which is funded in part by the National Science Foundation EPSCoR Research Infrastructure Improvement Cooperative Agreement \#1003907,
%the state of West Virginia (WVEPSCoR via the Higher Education Policy Commission) and WVU.
% we used Bowser not Spruce Knob...
Data for this project were collected using the facilities of the Green Bank Observatory and the Arecibo Observatory.
The National Radio Astronomy Observatory and Green Bank Observatory is a facility of the National Science Foundation operated under cooperative agreement by Associated Universities, Inc.
The Arecibo Observatory is a facility of the National Science Foundation operated under cooperative agreement by the University of Central Florida in alliance with Yang Enterprises, Inc.~and Universidad Metropolitana.

\appendix
%!TEX root = nanograv_11yr_bwm.tex

\section{An anomalous event in the 9-year analysis}
\label{sec:app.anom}

As stated in \autoref{sec:results.detect}, when analyzing the data from \citetalias{ng9} we find an anomalous GW memory detection.
The best fit sky position for this anomalous event is a region of low sensitivity (J2000, R.A. $4^\mathrm{h} 23^\mathrm{m}$, Dec.~$5^\circ 44^\prime$).
This sky location combined with the best fit polarization angle conspire to hide this event from nearly all of our pulsars.
To determine which pulsars are problematic we can perform a ``dropout analysis'' introduced in \citetalias{ng11.cw}.
For a dropout analysis we conduct the standard search with a few modifications.
First, we fix the parameters of the GW memory signal to match their best fit values from the previous analysis.
Next, we introduce a new parameter for each pulsar which acts as a switch.
Pulsars that are switched \textit{on} have the GW memory signal included in the likelihood calculation.
Pulsars that are switched \textit{off} do not.
From the posteriors of these parameters we are able to assess whether each pulsar prefers a noise only model or a noise plus GW memory model.
If a pulsar prefers to be in the \textit{off} state, then the noise only model is a better fit to its data.
\par

\begin{figure}
  \includegraphics[width=\columnwidth]{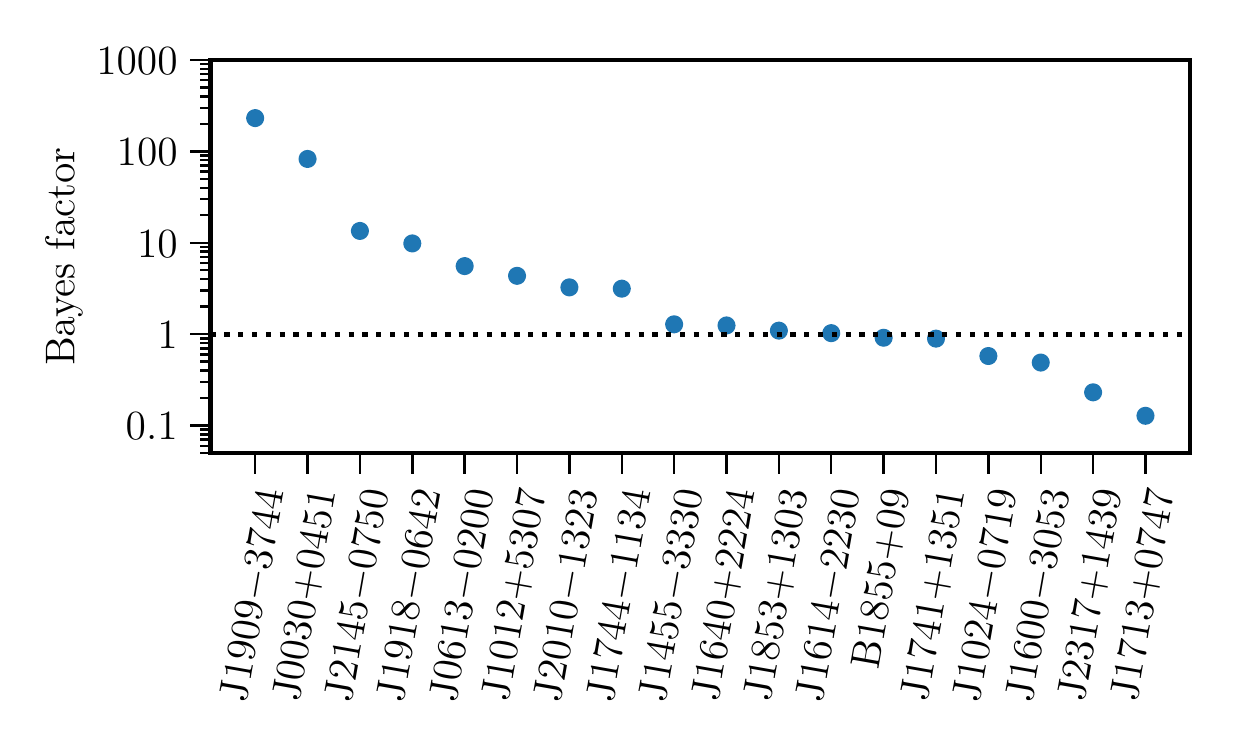}
  \caption{Results of dropout analysis for the anomalous event in the 9-year dataset.
  Bayes factors for the signal to be \textit{on} in each pulsar.
  The event is dominated by two pulsars, PSRs \psrAnomalyA~and \psrAnomalyB.
  This event was not found in the analysis of the full 11-year dataset indicating it is not a real GW event.
  }
  \label{fig:dropout}
\end{figure}

Bayes factors for \textit{on} versus \textit{off} for each pulsar are shown in \autoref{fig:dropout}.
PSRs \psrAnomalyA~and \psrAnomalyB~are the only two that show significant preference for the GW memory signal being turned \textit{on} with Bayes factors $\sim$100.
This tells us that the anomalous signal is isolated to these two pulsars and therefore unlikely to be a true GW memory signal.
We expect any real GW signal to appear significantly in several pulsars.
\par

Based on the 11-year single pulsar limits, the four most sensitive pulsars in the array to GW memory are PSRs \psrBestA, \psrAnomalyA, \psrBestB, and \psrBestC.
The first three were observed for the whole length of the data set, while \psrBestC~was added to the array in 2008.
It is worth noting that the three ``most \textit{off}'' pulsars are included in this list.
The presence of any \textit{off} pulsars should be a red flag for validating a detection.
If a pulsar is insensitive to a signal it should have no preference for \textit{on} or \textit{off}, therefore a Bayes factor $\sim1$.
The anomolous signal has a source orientation that already minimizes the response to most pulsars in the array.
The dropout analysis shows us that even with this supressed amplitude our most sensitive pulsars should still be able to see it, yet they do not.

While we are certain that this signal is not a GW, its origin remains a mystery.
Deeper investigation into the noise properties of the NANOGrav PTA is ongoing.
As the PTA continues to become more sensitive, new noise sources emerge which must be characterized and modeled.

%!TEX root = nanograv_11yr_bwm.tex

\section{Averaging over source orientation}
\label{sec:app.ave}
In order to have a fair comparison between the upper limits found from the Earth term, which were marginalized over source orientation, and the optimal oriented upper limits found from the pulsar term,
we must rescale the pulsar term limits accounting for the varying sensitivity depending on source orientation.
Most previous work followed \citet{vHl2010} and used the RMS average of the pulsar angular response function $\left\langle B^2 \right\rangle$ to account for source orientation.
Because the Earth term upper limit \textit{marginalizes} over source orientation, the fair comparison would do the same.
Here we analytically marginalize a pulsar's angular response $B_i(\hat{k}, \psi; \hat{n}_i)$, introduced in \autoref{eq:sigmod} over the GW source orientation.
\par

A pulsar's angular response to GW memory is given by \citet{ew1975} as (dropping the subscipt)
\begin{align}
	B(\hat{k}, \psi; \hat{n}) = B(\alpha, \beta) &= \frac{1}{2} \cos(2\beta) \left( 1 - \cos(\alpha) \right) \\
	\cos(\alpha) &= \hat{n} \cdot \hat{k}, \nonumber
\end{align}
where $\alpha$ is the angle between the line of sight to the pulsar $\hat{n}$ and the source $\hat{k}$,
and $\beta$ is the projected azimuthal angle between the source's principle polarization vector (defined by $\psi$) and $\hat{n}$ in the plane perpendicular to $\hat{k}$.
\par

For fixed $\hat{n}$ and $\hat{k}$ we can marginalize over the projected source polarization.
We integrate only the positive half cycle of $\beta$ using a uniform prior.
\begin{align}
	\frac{2}{\pi} \int_{-\pi/4}^{\pi/4} \dd\beta\, \cos(2\beta) = \frac{2}{\pi}
\end{align}
\par

Without loss of generality we can align the line of sight to the source with the $\hat{z}$-axis of our coordinate system.
Using spherical polar coordinates, $\hat{n} \cdot \hat{k} = \cos\theta$, where $\theta$ is co-latitude,
and $(\theta, \phi)$ is the source position on the sky.
We marginalize over source position using a uniform prior on the whole sphere of solid angle.
\begin{align}
	\frac{1}{4\pi} \int_{0}^{\pi} \dd\theta  \int_{0}^{2\pi} \sin\theta\, \dd\phi \, (1 - \cos\theta) = 1
\end{align}

Finally, we put it all together:
\begin{align}
	\frac{1}{2\pi^2} \int_{-\pi/4}^{\pi/4} \dd\beta\,\int_{0}^{\pi} \dd\theta  \int_{0}^{2\pi} \sin\theta\, \dd\phi\,  \left[ \frac{1}{2}(1 - \cos\theta) \cos(2\beta) \right] = \frac{1}{\pi}
\end{align}
We can rescale our single pulsar, optimally oriented upper limits by a factor of $\pi$ to account for non-uniform sensitivity.

\bibliographystyle{yahapj}
\bibliography{apjjabb,bib}

\end{document}